\documentclass[aps,prl,twocolumn,superscriptaddress,showpacs]{revtex4}
\usepackage{graphicx}
\begin{document}

\title{Nature of Orbital Ordering in La$_{0.5}$Sr$_{1.5}$MnO$_4$ Studied by Soft X-ray Linear Dichroism}
\author{D. J. Huang}
\affiliation{National Synchrotron Radiation Research Center, Hsinchu 30077, Taiwan}
\affiliation{Department of Electrophysics, National Chiao-Tung University, Hsinchu
300, Taiwan}
\author{W. B.
Wu} \affiliation{Department of Electrophysics, National Chiao-Tung University,
Hsinchu 300, Taiwan} \affiliation{National Synchrotron Radiation Research Center,
Hsinchu 30077, Taiwan}
\author{G. Y. Guo}
\affiliation{Department of Physics, National Taiwan University,
Taipei 106, Taiwan} \affiliation{National Synchrotron Radiation
Research Center, Hsinchu 30077, Taiwan}
\author{H.-J. Lin}
\author{T. Y. Hou}
\author{C. F. Chang}
\affiliation{National Synchrotron Radiation Research Center,
Hsinchu 30077, Taiwan}

\author{C. T. Chen}
\affiliation{National Synchrotron Radiation Research Center,
Hsinchu 30077, Taiwan} \affiliation{Department of Physics,
National Taiwan University, Taipei 106, Taiwan}
\author{A. Fujimori}
\affiliation{Department of Complexity Science and Department of Physics, University
of Tokyo, Tokyo 113-0033, Japan}
\author{T. Kimura}
\altaffiliation[Present address: ]{Los Alamos National Lab.,
MST-MISL, Mail Stop K774, Los Alamos, New Mexico 87545, U.
S.}\affiliation{Department of Applied Physics, University of
Tokyo, Tokyo 113-8656, Japan}
\author{H. B. Huang}
\affiliation{Department of Quantum Matters, ADSM, Hiroshima University,
Higashi-Hiroshima 739-8530, Japan}
\author{A. Tanaka}
\affiliation{Department of Quantum Matters, ADSM, Hiroshima University,
Higashi-Hiroshima 739-8530, Japan}
\author{T. Jo}
\affiliation{Department of Quantum Matters, ADSM, Hiroshima University,
Higashi-Hiroshima 739-8530, Japan}

\date{\today }
\begin{abstract}
We found that the conventional model of orbital ordering of
$3x^{2}-r^{2}$/$3y^{2}-r^{2}$ type in the $e_g$ states of
La$_{0.5}$Sr$_{1.5}$MnO$_4$ is incompatible with measurements of linear dichroism
in the Mn $2p$-edge x-ray absorption, whereas these $e_g$ states exhibit
predominantly cross-type orbital ordering of $x^{2}-z^{2}$/$y^{2}-z^{2}$. LDA+U
band-structure calculations reveal that such a cross-type orbital ordering results
from a combined effect of antiferromagnetic structure, Jahn-Teller distortion, and
on-site Coulomb interactions.

\end{abstract}

\pacs{75.47.Lx, 75.30.Mb, 71.28.+d, 78.70.Dm}

\maketitle Orbital ordering, which manifests itself in the spatial
distribution of the outermost valence electrons, is an important
topic in current research of transition-metal oxides, as the
magnetic and transport properties are closely related to the
orbital and charge degrees of freedom \cite{Tokura00}. In
particular, charge-orbital ordering of half-doped manganites has
attracted much attention
\cite{Goodenough55,Wollan55,Okimoto95,Schiffer95,Radaelli97a,Mizokawa97,Mori98}.
The mechanism of charge-orbital ordering is being hotly debated
\cite{Mutou99,Brink99,Khomskii00,Hotta01,Mutou01,Mahadevan01,Popovic02}.
To observe orbital ordering directly is a difficult task.
Experimental results of resonant x-ray scattering (RXS) at the Mn
$K$-edge in La$_{0.5}$Sr$_{1.5}$MnO$_4$ and LaMnO$_3$ show removal
of degeneracy between $4p_x$ and $4p_y$; these observations have
been argued to be direct evidence of orbital ordering
\cite{Murakami98a,Murakami98b}. However, the origin of RXS at Mn
$K$-edge is controversial. Orbital ordering in transition-metal
oxides is typically accompanied by Jahn-Teller lattice distortion.
Calculations based on a local-density approximation including
on-site Coulomb interactions (LDA+U) \cite{Elfimov99,Benedetti01}
and multiple scattering theory \cite{Benfatto99} indicate that RXS
measurements pertain mainly to Jahn-Teller distortion, instead of
directly observing $3d$ orbital ordering. Multiplet calculations
have shown that one can use linear dichroism (LD) in soft x-ray
absorption spectroscopy (XAS) to identify the orbital character of
$3d$ states in orbital-ordered manganites \cite{HBHuang00}.

Half-doped manganites such as La$_{0.5}$Sr$_{1.5}$MnO$_4$ exhibit CE-type
antiferromagnetic (AFM) ordering and charge-orbital ordering
\cite{Moritomo95,Sternlieb96,Murakami98a}. They have a zigzag magnetic structure in
which the magnetic moments of Mn on the chain form a ferromagnetic coupling, but
AFM coupling between the zigzag chains. Below the charge-ordering (CO) temperature
T$_{\rm CO}$=217~K, the valence of La$_{0.5}$Sr$_{1.5}$MnO$_4$ orders in an
alternating pattern with two distinct sites identified as Mn$^{3+}$ and Mn$^{4+}$
\cite{Moritomo95,Sternlieb96}. The $e_g$ electrons on the nominal Mn$^{3+}$ sites
of La$_{0.5}$Sr$_{1.5}$MnO$_4$ are believed to exhibit an orbital ordering of
$3x^{2}-r^{2}$/$3y^{2}-r^{2}$, in which occupied $d_{3x^{2}-r^{2}}$ and
$d_{3y^{2}-r^{2}}$ orbitals are alternately arranged at two sublattices in the $ab$
plane \cite{Mizokawa97}. However, $d_{3x^{2}-r^{2}}$ and $d_{x^{2}-z^{2}}$
($d_{3y^{2}-r^{2}}$ and $d_{y^{2}-z^{2}}$) orbitals might be mixed, because
orbitals of these two types have the same spatial symmetry with respect to the
MnO$_2$ plane. To clarify the nature of orbital ordering in
La$_{0.5}$Sr$_{1.5}$MnO$_4$ is essential to reveal the origin of orbital ordering
in half-doped manganites.

In this Letter, we present measurements of LD in Mn $2p$-edge XAS of
La$_{1-x}$Sr$_{1+x}$MnO$_4$. The LD measurements are compared with multiplet
calculations to unravel the orbital character of $e_g$ electrons in
La$_{0.5}$Sr$_{1.5}$MnO$_4$. We performed also LDA+U calculations to study the
orbital-ordering of this compound.

Single-crystalline samples of La$_{1-x}$Sr$_{1+x}$MnO$_4$ were
grown by the floating zone method \cite{Moritomo95}. Measurements
of x-ray diffraction at room temperature show that our samples are
of single phase. The major crystallographic difference between
crystals with different $x$'s is the $c$-axis length; this
decreases significantly from 13.04~${\rm \AA}$ for $x$=0 to
12.43~${\rm \AA}$ for $x$=0.5, whereas the $a$-axis length shows
only a weak $x$-dependence (3.81~${\rm \AA}$ for $x$=0 and
3.86~${\rm \AA}$ for $x$=0.5). This difference of the $c$-axis
length is attributed to a significantly decreased out-of-plane
Mn-O bond length with increasing Sr content.

XAS measurements on La$_{1-x}$Sr$_{1+x}$MnO$_4$ single crystals at various
temperatures were performed at the Dragon beamline of the National Synchrotron
Radiation Research Center in Taiwan. We recorded XAS spectra by collecting the
sample drain current. Crystals were freshly cleaved in an ultra-high vacuum at
90~K; the incident angle was $60^{\circ}$ from the sample surface normal and the
photon energy resolution was 0.2 eV. We rotated the sample about the direction of
incident photons to obtain LD spectra from which experimental artifacts related to
the difference in the optical path and to the probing area have been eliminated.
All measured XAS spectra referred to the \textbf{E} vector parallel to the \emph{c}
axis are shown with a correction for the geometry effect \cite{LD_set_up,Wu}.

\begin{figure}[tbp]
\includegraphics[width=8cm]{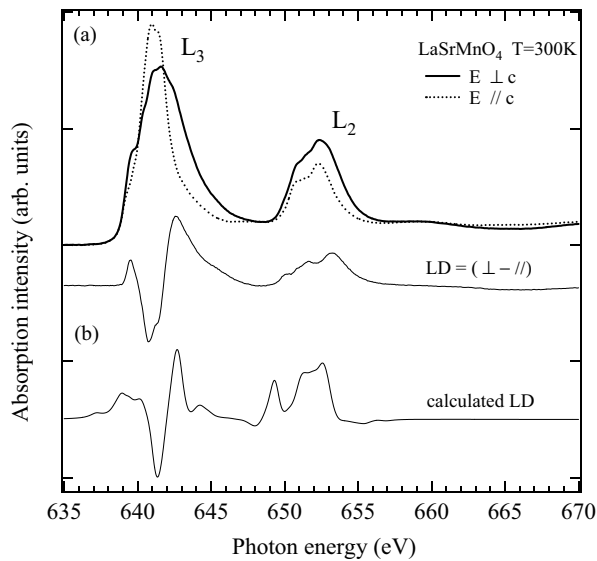}
\caption{\label{Fig_LD_undoped}(a) LD and polarization-dependent
XAS taken with \textbf{E} $\perp c$ (solid line) and \textbf{E}
$\parallel c$ (broken line) of LaSrMnO$_4$. (b) Calculated LD
spectrum of Mn$^{3+}$ ions with occupied $d_{3z^{2}-r^{2}}$
orbitals. The calculated LD is plotted on the same scale as the
measured one.}
\end{figure}

Multiplet calculations have shown that one can use LD in $L$-edge
XAS to characterize the $3d$ orbital character \cite{HBHuang00}.
LD in XAS is defined as the difference between XAS spectra taken
with the \textbf{E} vector of photons perpendicular and parallel
to the crystal $c$-axis. To verify experimentally such a
capability of LD, we measured the LD in Mn $L_{2,3}$-edge XAS of
LaSrMnO$_4$, which is expected to exhibit $3z^{2}-r^{2}$
"ferro-orbital" ordering. Figure \ref{Fig_LD_undoped}(a) shows our
measurements of polarization-dependent XAS and LD on LaSrMnO$_4$.
Most features in the measured LD at Mn L-edge are reproduced by
multiplet calculations for Mn$^{3+}$ ions with occupied
$d_{3z^{2}-r^{2}}$ orbitals, as shown in Fig.
\ref{Fig_LD_undoped}(b) \cite{HBHuang00,Multiplet}, revealing that
LD in L-edge XAS is an effective means to examine the orbital
character of $3d$ electronic states in an orbital-ordered
compound.

\begin{figure}[tbp]
\includegraphics[width=8cm]{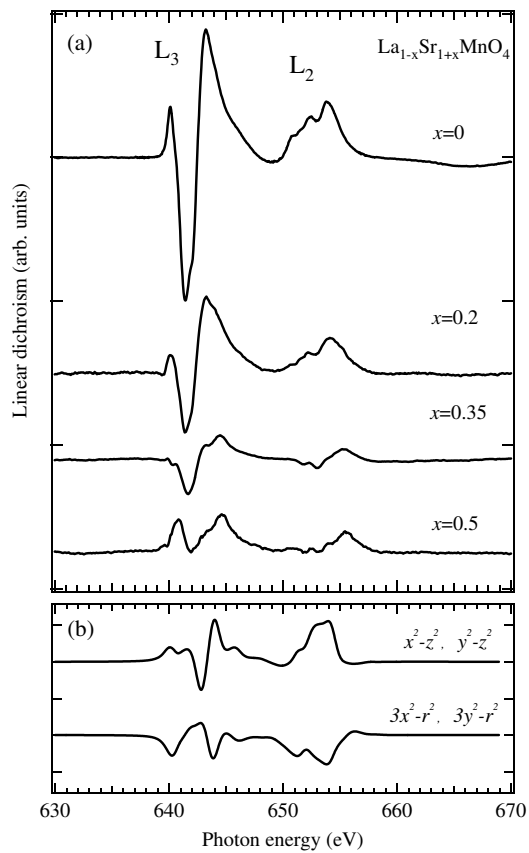}
\caption{\label{Fig_LD_doping}(a) LD in Mn $L_{2,3}$-edge XAS of
La$_{1-x}$Sr$_{1+x}$MnO$_4$ with varied doping. Linear-dichroism
spectra were derived from XAS normalized to the same peak
intensity at Mn $L_{3}$-edge and measured at 300~K for $x\leq
0.35$ and 150~K for $x=0.5$. (b) Calculated LD spectra of
Mn$^{3+}$ ions with $d_{x^{2}-z^{2}}$/$d_{y^{2}-z^{2}}$ and
$d_{3x^{2}-r^{2}}$/$d_{3y^{2}-r^{2}}$ orbitals occupied.}
\end{figure}

\begin{figure}[tbp]
\includegraphics[width=8cm]{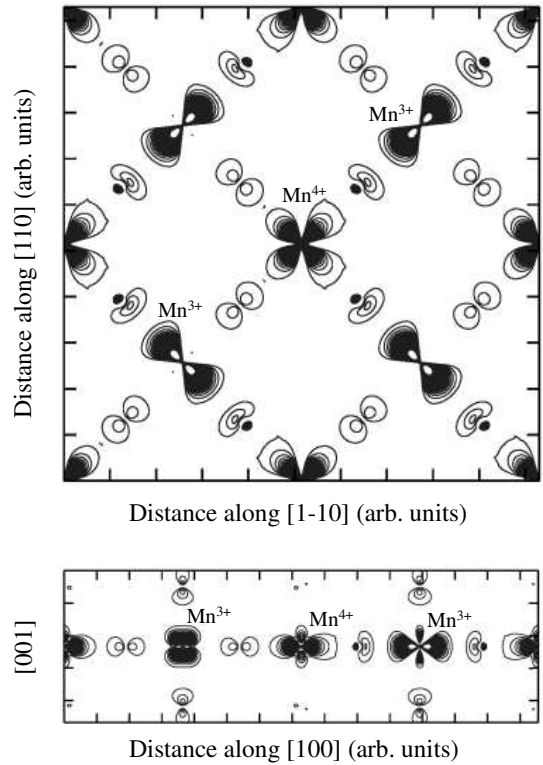}
\caption{\label{Fig_Contour} Charge-density contours corresponding
to the $e_g$ valence bands of La$_{0.5}$Sr$_{1.5}$MnO$_4$. Upper
panel: Charge-density contours in the $ab$ plane. Lower panel:
Charge-density contours in the $ac$ plane along [100] direction.}
\end{figure}

We measured also the LD in XAS on La$_{1-x}$Sr$_{1+x}$MnO$_4$ with
varied doping to clarify further the origin of the LD signal, as
shown in Fig. \ref{Fig_LD_doping}(a). Because the Jahn-Teller
effect on the Mn$^{4+}$ ions is insignificant
\cite{Mahadevan01,Popovic02}, the contribution of $3d$ orbitals of
these ions to LD is much smaller than that of Mn$^{3+}$ ions. With
increasing doping, the proportion of Mn$^{3+}$ ions decreases; the
LD magnitude of doped La$_{1-x}$Sr$_{1+x}$MnO$_4$ diminishes. Note
that the magnitude of LD decreases more rapidly than that from a
simple picture of Mn$^{3+}$-Mn$^{4+}$ dilution. In particular, the
LD magnitude of La$_{0.5}$Sr$_{1.5}$MnO$_4$ is $\sim 1/4$ that
observed for LaSrMnO$_4$; its sign at the $L_2$-edge is the same
as that of LaSrMnO$_4$. To identify the orbital character of the
occupied $e_g$ states, by using a model of MnO$_6$ cluster based
on configuration interaction \cite{Elp99}, we calculated LD
spectra of Mn$^{3+}$ with $d_{x^{2}-z^{2}}$/$d_{y^{2}-z^{2}}$ and
$d_{3x^{2}-r^{2}}$/$d_{3y^{2}-r^{2}}$ orbitals occupied, as shown
in Fig. \ref{Fig_LD_doping}(b) \cite{CI}. Overall the calculated
LD of occupied in-plane orbitals such as $d_{3x^{2}-r^{2}}$ and
$d_{3y^{2}-r^{2}}$ is with sign reversed to that of out-of-plane
orbitals such as $d_{3z^{2}-r^{2}}$, $d_{x^{2}-z^{2}}$, and
$d_{y^{2}-z^{2}}$. Surprisingly the conventional orbital ordering
model of $3x^{2}-r^{2}$/$3y^{2}-r^{2}$ type is incompatible with
LD measurements. The calculated LD of
$3x^{2}-r^{2}$/$3y^{2}-r^{2}$-type orbital ordering is with sign
reversed to that of measured LD from La$_{0.5}$Sr$_{1.5}$MnO$_4$.
One might suspect this inconsistency could result from anisotropic
$e_g$ charge distribution on the Mn$^{4+}$ sites. If so, only
$e_g$ charge with $d_{3z^{2}-r^{2}}$ or
$d_{x^{2}-z^{2}}$/$d_{y^{2}-z^{2}}$ polarization transferred from
Mn$^{3+}$ to Mn$^{4+}$ could give rise to a LD similar to the
measurement. However, even in the most unfavorable case, that is,
even if the transferred $e_g$ charge on the Mn$^{4+}$ site were
maximum (leading to equal charges on both Mn sites) and fully
($3z^{2}-r^{2}$)-polarized, only half of the observed LD could be
accounted for. As shown later (see the lower panel of Fig.
\ref{Fig_Contour}), such transferred $e_g$ charges indeed have a
small in-plane polarization. This anisotropy gives opposite
contributions to LD with respect to the measurement; the
inconsistency can not be reconciled even if the anisotropic charge
distribution of Mn$^{4+}$ was taken into account.

Furthermore, the lineshape of the measured LD spectrum for $x=0.5$
is similar to those from calculations for Mn$^{3+}$ with occupied
$d_{3z^{2}-r^{2}}$ or $d_{x^{2}-z^{2}}$/$d_{y^{2}-z^{2}}$
orbitals, implying that La$_{0.5}$Sr$_{1.5}$MnO$_4$ has an orbital
polarization of strong $z$ character, \emph{e.g.},
$d_{3z^{2}-r^{2}}$ or $d_{x^{2}-z^{2}}$/$d_{y^{2}-z^{2}}$. If
La$_{0.5}$Sr$_{1.5}$MnO$_4$ exhibited $3z^{2}-r^{2}$ orbital
ordering, all Mn$^{3+}$ sites, \emph{i.e.,} half of all Mn atoms,
would contribute to LD and its magnitude at Mn $L_2$-edge would be
half of that observed in LaSrMnO$_4$, in contrast to the
measurements. If La$_{0.5}$Sr$_{1.5}$MnO$_4$ exhibits
$x^{2}-z^{2}$/$y^{2}-z^{2}$ orbital ordering, by choosing LD as
the difference in XAS spectra taken with the \textbf{E} vector
parallel to $x$ and $z$ axes, we observe essentially linear
dichroism resulting only from the sublattice with occupied
$d_{y^{2}-z^{2}}$. In other words, only half of Mn$^{3+}$ sites
contribute to LD; one quarter of Mn atoms contribute to LD,
consistent with the measurements. Our LD measurements thus suggest
that orbital ordering of the $e_g$ states on the Mn site in
La$_{0.5}$Sr$_{1.5}$MnO$_4$ is dominated by
$x^{2}-z^{2}$/$y^{2}-z^{2}$ type.

To further study orbital ordering in La$_{0.5}$Sr$_{1.5}$MnO$_4$,
we performed LDA+U calculations using the full-potential
linearized augmented-plane-wave method on
La$_{0.5}$Sr$_{1.5}$MnO$_4$ in CE-type AFM structure with $U$ and
$J$ equal to 8 and 0.88 eV for Mn $3d$ electrons, respectively
\cite{Elfimov99}. Details of the calculations will be described
elsewhere \cite{Guo}. Mahadevan \emph{et al.} \cite{Mahadevan01}
found that the breathing-type Jahn-Teller distortion of
La$_{0.5}$Sr$_{1.5}$MnO$_4$ suggested by Sternlieb \emph{et al.}
\cite{Sternlieb96} is not energetically favorable, and proposed a
shear-type Jahn-Teller distortion in which the Mn-O length is
elongated alternately along the $x$ and $y$ directions.
Measurements of x-ray scattering also indicate a shear-type
distortion on Mn-O octahedra \cite{Larochelle01}, rather than a
breathing-type distortion. Consistent with previous band-structure
calculations \cite{Mahadevan01}, our LDA+U calculations show also
that La$_{0.5}$Sr$_{1.5}$MnO$_4$ without Jahn-Teller distortion is
unstable against a shear-type Jahn-Teller distortion. With a
shear-type Jahn-Teller distortion of 0.08-{\AA} in-plane O
displacement \cite{distortion}, LDA+U calculations give rise to an
orbital ordering dominated by $x^{2}-z^{2}$/$y^{2}-z^{2}$ on the
Mn$^{3+}$ sites of La$_{0.5}$Sr$_{1.5}$MnO$_4$, as shown in Fig.
\ref{Fig_Contour}, which displays charge-density contours
corresponding to the $e_g$ dominated valence bands. Interestingly
we found also that La$_{0.5}$Sr$_{1.5}$MnO$_4$ would exhibit
$3x^{2}-r^{2}$/$3y^{2}-r^{2}$ orbital ordering if the on-site
Coulomb interactions were not explicitly included, in agreement
with previous LDA calculations \cite{Mahadevan01}. Our results
suggest that charge and orbital ordering  can be well described if
the on-site Coulomb interactions of $3d$ electrons are properly
taken into account, as in the LDA+U or Hartree-Fock calculations.
Such a cross-type orbital ordering results from a combined effect
of AFM structure, Jahn-Teller distortion, and the on-site Coulomb
interactions of $3d$ electrons.

\begin{figure}[tbp]
\includegraphics[width=8cm]{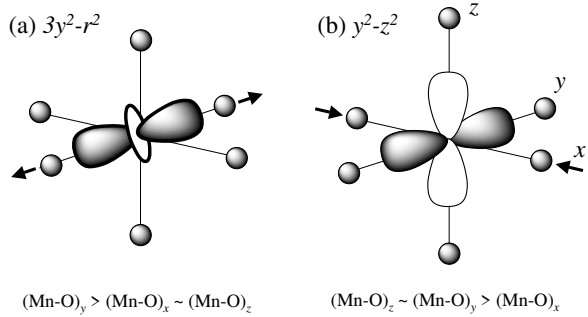}
\caption{\label{Fig_OO} View of $d_{3y^{2}-r^{2}}$ and
$d_{x^{2}-z^{2}}$ orbitals on the Mn$^{3+}$ sites with different
Jahn-Teller distortions. (a) and (b) show the Mn-O length
elongated along the $y$ direction and contracted along the $x$
direction, respectively. Filled circles denote O atoms in which
$2p$ orbitals are omitted for clarity.}
\end{figure}

The existence of orbital ordering of cross-type
$x^{2}-z^{2}$/$y^{2}-z^{2}$ can be understood within the framework
of crystal field effect with lattice distortion taken into
account. On the Mn$^{3+}$ sites of a cubic perovskite, $e_g$
orbitals of $3y^{2}-r^{2}$ ($3x^{2}-r^{2}$) symmetry are
preferentially occupied if the Mn-O length is elongated along the
$y$ ($x$) direction; $y^{2}-z^{2}$ ($x^{2}-z^{2}$) orbitals are
occupied if the Mn-O length is contracted along the $x$ ($y$)
direction, as shown in Fig. \ref{Fig_OO}. For example, in CE-type
charge-orbital-ordered half-doped manganites of cubic perovskite
such as La$_{0.5}$Ca$_{0.5}$MnO$_{3}$, the Mn$^{3+}$ site exhibits
a large Jahn-Teller distortion, in which the Mn-O length is
elongated alternately along the $x$ and $y$ directions (two long
bonds of 2.06~{\AA} along the zigzag chain and four short bonds of
1.92~{\AA}) \cite{Radaelli97b}, producing
$3x^{2}-r^{2}$/$3y^{2}-r^{2}$ orbital ordering. As for
La$_{0.5}$Sr$_{1.5}$MnO$_4$, the shear-type distortion leads
effectively to alternate contractions of along the $x$ and $y$
directions in La$_{0.5}$Sr$_{1.5}$MnO$_4$, because the longer
in-plane Mn-O length (2.00~{\AA}) is close to the out-of-plane
Mn-O length (1.98~{\AA}), while  the shorter in-plane Mn-O length
is 1.84~{\AA}. Orbital ordering of $x^{2}-z^{2}$/$y^{2}-z^{2}$ is
expected to be energetically more favorable than that of
$3x^{2}-r^{2}$/$3y^{2}-r^{2}$. Note that small tetragonal
distortions with $c/a$=0.98 and $c/a$=1.04 in strained thin films
of La$_{0.5}$Sr$_{0.5}$MnO$_3$ can result in ferro-orbital
ordering of $x^{2}-y^{2}$ and $3z^{2}-r^{2}$, respectively
\cite{Konishi99,Fang00}.

\begin{figure}[tbp]
\includegraphics[width=8cm]{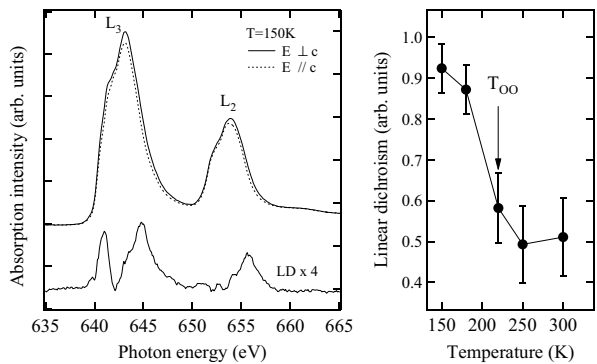}
\caption{\label{Fig_LD_temp}Left panel: XAS and LD spectra of
La$_{0.5}$Sr$_{1.5}$MnO$_4$ recorded at temperatures 150~K. Right
panel: temperature dependence of relative LD in XAS measured at
temperatures about T$_{\rm CO}$. The solid line is for visual
guidance. Error bars reflect uncertainties from different sets of
data.}
\end{figure}

In addition, measurements of temperature-dependent LD of
La$_{0.5}$Sr$_{1.5}$MnO$_4$ about T$_{\rm CO}$ provide us with
further evidence that LD in XAS reflects the nature of orbital
ordering. LD measurements with photon energy of 645~eV show that
the LD decreases greatly as the temperature crosses T$_{\rm CO}$,
as shown in Fig. \ref{Fig_LD_temp}, indicating that orbital
ordering of La$_{0.5}$Sr$_{1.5}$MnO$_4$ follows a similar
temperature-dependent trend of charge ordering \cite{Sternlieb96}
and Jahn-Teller distortion \cite{Murakami98a}. To confirm this,
more detailed temperature dependent studies would be necessary.

To conclude, we demonstrate that LD in Mn $2p$ XAS is a powerful
method to test the validity of models for orbital ordering in
transition-metal oxides. With LD measurements, we inferred that
orbital ordering of the Mn $e_g$ electrons in
La$_{0.5}$Sr$_{1.5}$MnO$_4$ is dominated by
$x^{2}-z^{2}$/$y^{2}-z^{2}$ type, as corroborated by our LDA+U
calculations. Orbital ordering of Mn $e_g$ electrons in
La$_{0.5}$Sr$_{1.5}$MnO$_4$ results from a combined effect of
antiferromagnetic structure, Jahn-Teller distortion, and on-site
Coulomb interactions. In principle, one can directly observe both
orbital ordering and Jahn-Teller ordering in manganites by using
resonant x-ray scattering at Mn L$_{2,3}$-edges
\cite{Castleton00}.

\begin{acknowledgements}
We thank Y. Tokura, T. Mizokawa, P. Mahadevan, and C. H. Chen for
valuable discussions. We also thank L. H. Tjeng for loaning his
XAS chamber. This work was supported in part by the National
Science Council of Taiwan and by a Grant-in-Aid for Scientific
Research in Priority Area "Novel Quantum Phenomena in
Transition-Metal Oxides" from the Ministry of Education, Culture,
Sports, Science and Technology of Japan.
\end{acknowledgements}


\begin{thebibliography}{99}
\bibitem{LD_set_up} XAS intensity $I_{\parallel}$ for \textbf{E}~${\parallel}~c$ is
deduced from $I_{\parallel}=\frac{4}{3}(I-\frac{1}{4}I_{\perp})$, in which
$I_{\perp}$ and $I$ are XAS intensities measured with \textbf{E}~${\perp}~c$ and
with \textbf{E} in the plane defined by the $c$ axis and the direction of incident
radiation, respectively.
\bibitem{Multiplet} We used $10Dq$ = 2.2~eV,
$\Delta_{i}$ = 0.1~eV, and $\Delta_{f}$ = 0.5~eV. Other
paparameters are the same as those in Ref. \cite{HBHuang00}.
\bibitem{CI} We used $U$ = 8.0 eV, $10Dq$ = 2.0
eV, charge-transfer energy $\Delta$ = 2.1 eV, and $pd\sigma$ = 2.1
eV.
\bibitem{distortion} The longer and shorter in-plane Mn-O lengthes are, respectively,
2.00~{\AA} and 1.84~{\AA}, while the out-of-plane Mn-O length is
1.98~{\AA}.
\end{thebibliography}
\end{document}